\documentclass[twocolumn,prl,showpacs,superscriptaddress,showpacs]{revtex4-1}
\usepackage{graphicx}
\usepackage{amssymb,amsmath}
\usepackage{bm}
\usepackage{dcolumn}
\usepackage{subfigure}
\usepackage{float}
\usepackage{url}
\usepackage{color}
\usepackage{txfonts}
\usepackage[driverfallback=dvipdfm]{hyperref}
\hypersetup{pdfpagemode=FullScreen,colorlinks=true,breaklinks,urlcolor=blue,linkcolor=blue,citecolor=blue}
\usepackage{natbib}

\begin{document}

\title{Symmetry Protected Dynamical Symmetry in the Generalized Hubbard Models}

\author{Jinlong Yu}
\affiliation{Institute for Advanced Study, Tsinghua University, Beijing, 100084, China}
\author{Ning Sun}
\affiliation{Institute for Advanced Study, Tsinghua University, Beijing, 100084, China}
\author{Hui Zhai}
\affiliation{Institute for Advanced Study, Tsinghua University, Beijing, 100084, China}
\affiliation{Collaborative Innovation Center of Quantum Matter, Beijing, 100084, China}

\date{\today }

\begin{abstract}
In this letter we present a theorem on the dynamics of the generalized Hubbard models. This theorem shows that the symmetry of the single particle Hamiltonian can protect a kind of dynamical symmetry driven by the interactions. Here the dynamical symmetry refers to that the time evolution of certain observables are symmetric between the repulsive and attractive Hubbard models. We demonstrate our theorem with three different examples in which the symmetry involves bipartite lattice symmetry, reflection symmetry and translation symmetry, respectively. Each of these examples relates to one recent cold atom experiment on the dynamics in the optical lattices where such a dynamical symmetry is manifested. These experiments include expansion dynamics of cold atoms, chirality of atomic motion within a synthetic magnetic field and melting of charge-density-wave order. Therefore, our theorem provides a unified view of these seemingly disparate phenomena.
 
\end{abstract}

\maketitle

The Hubbard model lies at the heart of studying the strongly correlated quantum matters~\cite{Hubbard1964,Fisher1989,HubbardModelReview1996,SachdevBook1999,KeimerReview2015}. It describes either fermions or bosons hopping in a lattice with short-range interactions ~\cite{Hubbard1964,Fisher1989}. Normally the Hubbard model considers a single band situation, and its Hamiltonian is written as 
\begin{equation}
\hat{H}=\hat{H}_0+\hat{V},
\end{equation}
where $\hat{H}_0$ is the single-particle term and $\hat{V}$ represents the on-site interaction between particles. For spinless bosons, 
\begin{equation}
\hat{V}=U\sum_{i}n_i (n_i-1), \label{Vboson}
\end{equation}
where $n_i$ is the density of bosons at site $i$; for spin-1/2 fermions, 
\begin{equation}
\hat{V}=U\sum_{i}n_{i\uparrow}n_{i\downarrow}, \label{Vfermion}
\end{equation}
where $n_{i\sigma}$ ($\sigma=\uparrow,\downarrow$) is the density of fermions with spin $\sigma$ at site $i$. These two cases are called bosonic and fermionic Hubbard models, respectively. Here $U$ represents the interaction strength, and $U>0$ ($U<0$) means repulsive (attractive) interaction. For the simplest situation, the single-particle Hamiltonian $\hat{H}_0$ only contains the (real-valued) nearest neighboring hopping terms. In more involved settings, it can also contain terms such as the periodic modulation of the on-site energy~\cite{Munich2015}, and the gauge fields can also add extra phases into the hopping coefficients~\cite{Bloch2013,Ketterle2013}. Here we term these interacting models with different $\hat{H}_0$ as the generalized Hubbard models. 

In the past decades, the Hubbard model is also a central topic for the cold atom quantum simulations~\cite{JakschZoller2005,BlochReview2008,BlochReview2012,LewensteinBook2012}. The reasons are at least two folds. Firstly, by loading ultracold bosons or fermions into optical lattices, the system is a faithful representation of the bosonic or fermionic Hubbard model, because the multi-bands effect and the longer rang interaction are sufficiently weak that can be safely ignored~\cite{BlochReview2008}. Secondly, these cold atom systems are particularly suitable for studying quantum dynamics~\cite{AltmanReview2015} in these strongly correlated systems, which is less studied in previous investigations in the content of condensed matter systems. For instance, one can first prepare this many-body system in a certain initial state, and experimentally observe the time evolution of this state. In the past decade, quite a few experiments have carried such investigations. Here we briefly review three of them:

\textbf{Munich 2012}: In this experiment from the Munich group, they first prepare the Fermi gas in a band insulator state in the presence of a harmonic trap, and then they turn off the harmonic trap and let the gas expand in a uniform three-dimensional cubic lattice~\cite{Munich2012}. The dramatic finding is that the expansion dynamics is identical between two systems with $+U$ and $-U$. In a related earlier experiment, the same group also found that a Fermi gas expands (instead of shrinks) when interaction becomes attractive, which is quite counter intuitive~\cite{Munich2010}. 

\textbf{Munich 2015}: Motivated by many-body localization, the Munich group investigates the relaxation of a charge-density-wave (CDW) state of fermions in the presence of an incommensurate lattice potential~\cite{Munich2015}. They observe how the CDW order evolves in time and saturates at longer times. As a side result, they also find that the dynamics of this CDW order is symmetric between positive and negative $U$. 

\textbf{Harvard 2017}: The Harvard group realized a two-leg Harper-Hofstadter model, in which there exists a uniform synthetic magnetic flux though each plaquette~\cite{Harvard2017}. They focus on studying the chirality in this model by loading one or two bosons into the ladder. Here the chirality means that the wave function is more concentrated in the upper ladder when atoms move to left (right), while it is more concentrated in the lower ladder when atoms move to the right (left). Such a chiral motion has been observed for the single particle case. However, considering the two-atom case with certain initial state, surprisingly they find that the chirality vanishes if no interaction is applied, and the chirality is induced when the interaction is turned on.

One common feature of all these three experiments is that the time evolution of certain observable is symmetric between repulsive and attractive interaction models. Following Ref.~\cite{Munich2012}, we term this symmetry as a kind of ``dynamical symmetry". The main result of this letter is to present the following theorem. It shows that the existence of a symmetry for the single particle Hamiltonian [Eq. (\ref{Eq:S_H})] is a key to guaranteeing this dynamical symmetry. Therefore, we term the phenomenon described by this theorem as ``symmetry protected dynamical symmetry"~\cite{footnote}. The significance of this theorem is that it shows that the symmetry of the single particle Hamiltonian can impose a strong constraint on the dynamics induced by the interactions. We will show that all the above three experimental observations can be understood as special examples of this theorem. 

\textbf{Theorem.} For the Hamiltonian $\hat{H}=\hat{H}_0+\hat{V}$, if we can find an antiunitary operator $\hat{S} = \hat{R}\hat{W}$, where $\hat{R}$ is the (antiunitary) time-reversal operator and $\hat{W}$ is a unitary operator that satisfies the following conditions: 

(i) $\hat{S}$ anticommutes with $\hat{H}_0$ and commutes with $\hat{V}$, i.e. 
\begin{equation} \label{Eq:S_H}
  \{ \hat{S},\hat{H}_0\}  = 0, \quad [\hat{S},\hat{V}] = 0;
\end{equation}

(ii) The initial state $\left| {{\Psi _0}} \right\rangle $ only acquires a global phase factor under $\hat{S}$, i.e. 
\begin{equation} \label{Eq:psi_0}
  \hat{S}^{ - 1}\left| {{\Psi _0}} \right\rangle  = {e^{i\chi }}\left| {{\Psi _0}} \right\rangle;
\end{equation}

(iii) We consider a given Hermitian operator $\hat{O}$ that is even or odd under symmetry operation by $\hat{S}$, i.e. 
\begin{equation} \label{Eq:SOS}
  \hat{S}^{ - 1}\hat{O}\hat{S} =  \pm \hat{O},
\end{equation}
then we can conclude
\begin{equation} \label{Eq:O_t_pmU}
  {\left\langle {O(t)} \right\rangle _ {+U} } =  \pm {\left\langle {O(t)} \right\rangle _ {-U} }, 
\end{equation} 
where $\left\langle {O(t)} \right\rangle _ {\pm U}$ denotes the expectation value of $\hat{O}$ under the wave function $|\Psi(t)\rangle=e^{i\hat{H}t}|\Psi_0\rangle$ with interaction strength $\pm U$ in $\hat{H}$, respectively.

\textit{Proof of the Theorem.} The proof of this theorem is straightforward. First, we use condition (i) and obtain
\begin{equation}
  \begin{aligned}
  \hat{S}^{ - 1}{e^{ - i(\hat{H}_0 + \hat{V})t}}\hat{S} =& \exp \left[ {i\hat{S}^{ - 1}(\hat{H}_0 + \hat{V})\hat{S}t} \right] \hfill \\
   =& \exp \left[ { - i(\hat{H}_0 - \hat{V})t} \right].  \label{shs} 
\end{aligned} 
\end{equation}
Here, $\hat{R}^{-1}i \hat{R} = - i$ is used in the first line. 
Then, with Eq. (\ref{shs}) and using conditions (ii) and (iii), we obtain
\begin{equation}
	\begin{aligned}
  &{\left\langle \hat{O}(t) \right\rangle _{ + U}} = \left\langle {{\Psi _0}} \right|{e^{i({\hat{H}_0} + \hat{V})t}}\hat{O}{e^{ - i(\hat{H}_0 + \hat{V})t}}\left| {{\Psi _0}} \right\rangle  \hfill \\
   &= \left\langle {{\Psi _0}} \right|\hat{S}e^{i(\hat{H}_0 - \hat{V})t}\left( {{\hat{S}^{ - 1}}\hat{O}\hat{S}} \right)e^{-i(\hat{H}_0 -\hat{V})t}{\hat{S}^{ - 1}}\left| {{\Psi _0}} \right\rangle  \hfill \\
   &= \left\langle {{\Psi _0}} \right|{e^{ - i\chi }}\left( {{e^{i(\hat{H}_0 - \hat{V})t}}} \right)\left( { \pm \hat{O}} \right)\left( {{e^{ - i(\hat{H}_0 - \hat{V})t}}} \right){e^{i\chi }}\left| {{\Psi _0}} \right\rangle  \hfill \\
   &=  \pm {\left\langle {O(t)} \right\rangle _{-U}}. 
\end{aligned} 
\end{equation}
Hence the theorem is proved. 

Here we should remark that our theorem, as well as the examples below, work equally well for both the bosonic and fermionic Hubbard models with interaction terms Eq. (\ref{Vboson}) and Eq. (\ref{Vfermion}), respectively. Hereafter different examples have different $\hat{H}_0$ and we use $\hat{c}_{i\sigma}$ to denote the annihilation operator for either a boson or a fermion in site $i$ with spin $\sigma$. (For spinless bosons, the $\sigma$ index can be ignored.)

\begin{figure}[t]
  \includegraphics[width=\columnwidth]{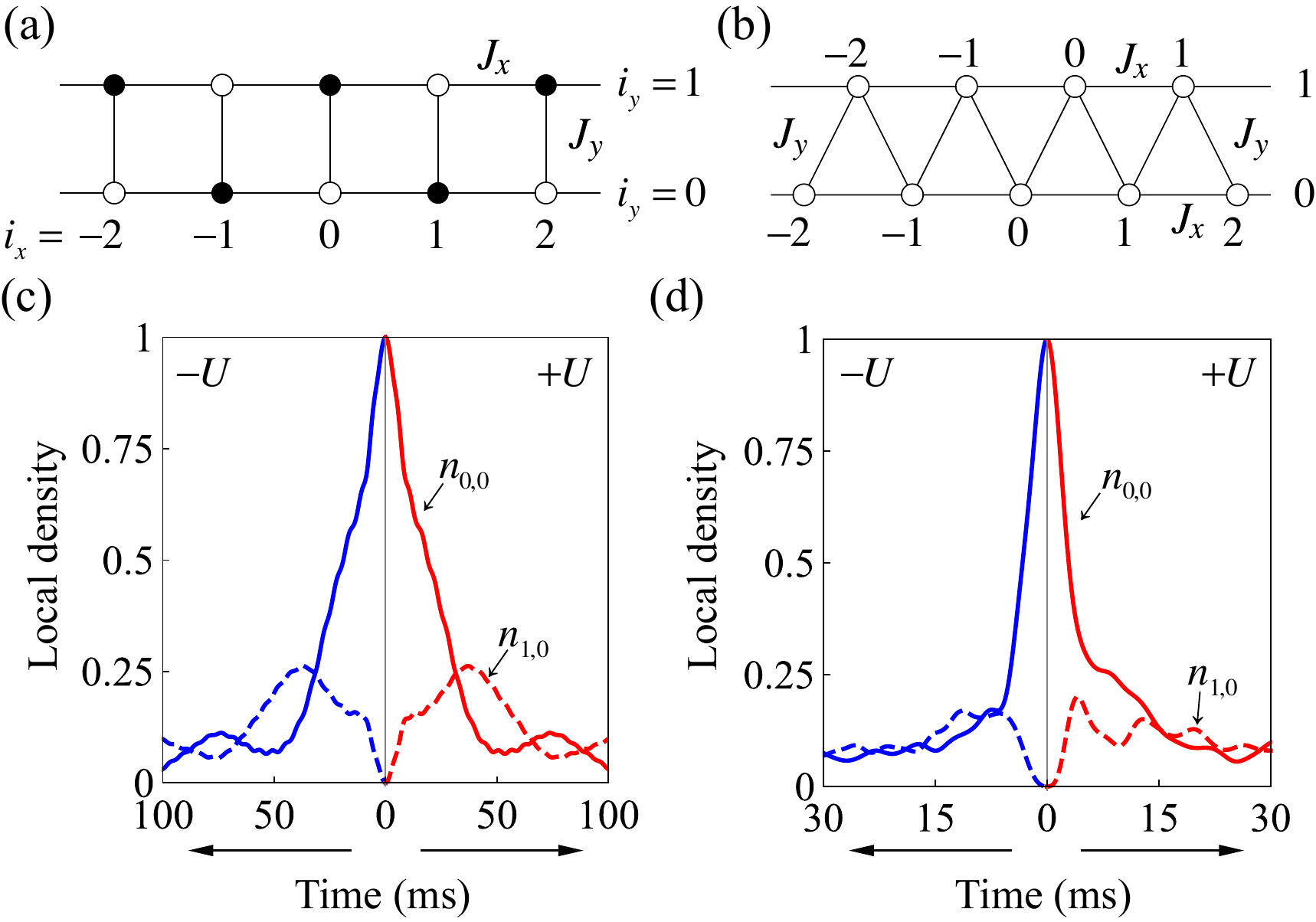}\\
  \caption{The presence (absence) of dynamical symmetry in the square (triangular) ladder. (a) and (b) shows the ladder configuration. (c) The time evolution of the local densities of the square ladder with attractive (left half) and repulsive (right half) interactions. The initial state is a two-boson state located at the central rung: $|\Psi_0\rangle = \hat{c}^\dag_{0,0}\hat{c}^\dag_{0,1}|0\rangle$. 
  Here we take $(J_x, \,J_y,\, U)=2\pi\times(11,\, 34, \,131)\,\text{Hz}$. (d) The same as (c), except that we consider the triangular ladder case.}\label{Fig1}
\end{figure}

\textit{Example 1:} We consider particles hop with nearest neighboring hopping only, in which 
\begin{equation}
H_0=-J\sum\limits_{\langle ij\rangle,\sigma}c^\dag_{i\sigma}c_{j\sigma},
\end{equation}
where $J$ is the hopping amplitude. This is the simplest case, and it well explains the \textbf{Munich 2012} experiment~\cite{Munich2012}. In fact, similar discussion specifically made to this model has been presented in Ref.~\cite{Munich2012}. Nevertheless, we view it as one application of our theorem and review it here for comprehensiveness for general readers. 

Obviously, this $\hat{H}_0$ is invariant under time-reversal operation. If the lattice is a bipartite lattice containing $A$ and $B$ sublattices, say, a square lattice, we have a symmetry operator $\hat{W}$ defined as
\begin{align}
&\hat{W}^{-1}\hat{c}_{i\sigma}\hat{W}=-\hat{c}_{i\sigma}, \   \  \text{if} \   \  i\in A, \label{W1a} \\
&\hat{W}^{-1}\hat{c}_{i\sigma}\hat{W}=\hat{c}_{i\sigma}, \   \  \text{if} \   \  i\in B. \label{W1b}
\end{align} 
Because for a bipartite lattice, hopping only takes place between $A$ and $B$ sublattices, it is easy to show that, with this choice of $\hat{W}$, $\hat{S}^{-1}\hat{H}_0\hat{S}=-\hat{H}_0$. And it is also easy to show that this transformation $\hat{S}$ leaves $\hat{V}$ invariant. Thus, we have found an operation $\hat{S}$ satisfying condition (i) of our theorem. It is also easy to see that, when the initial state is chosen as a band insulator, it is invariant under $\hat{S}$ and condition (ii) is satisfied; and when the observable is density operator $\hat{n}_{i\sigma}$, it satisfies condition (iii) with a plus sign. Thus, our theorem applies. We should also remark, here the bipartite lattice geometry plays a crucial role. If the lattice is not bipartite, say, a triangular lattice, we can not find such a $\hat{W}$.  

In Fig.~\ref{Fig1}, we numerically demonstrate this 
statement by loading two interacting bosons into two kinds of ladders with different geometries.
The initial state is chosen as two bosons placed at two nearest neighboring sites. We find that the time-dependent local densities at different sites obey this dynamical symmetry when the lattice is a square lattice [Fig.~\ref{Fig1}(c)], and do not obey this dynamical symmetry when the lattice is a triangular one [Fig.~\ref{Fig1}(d)]. 

\begin{figure}[btph]
  \includegraphics[width=\columnwidth]{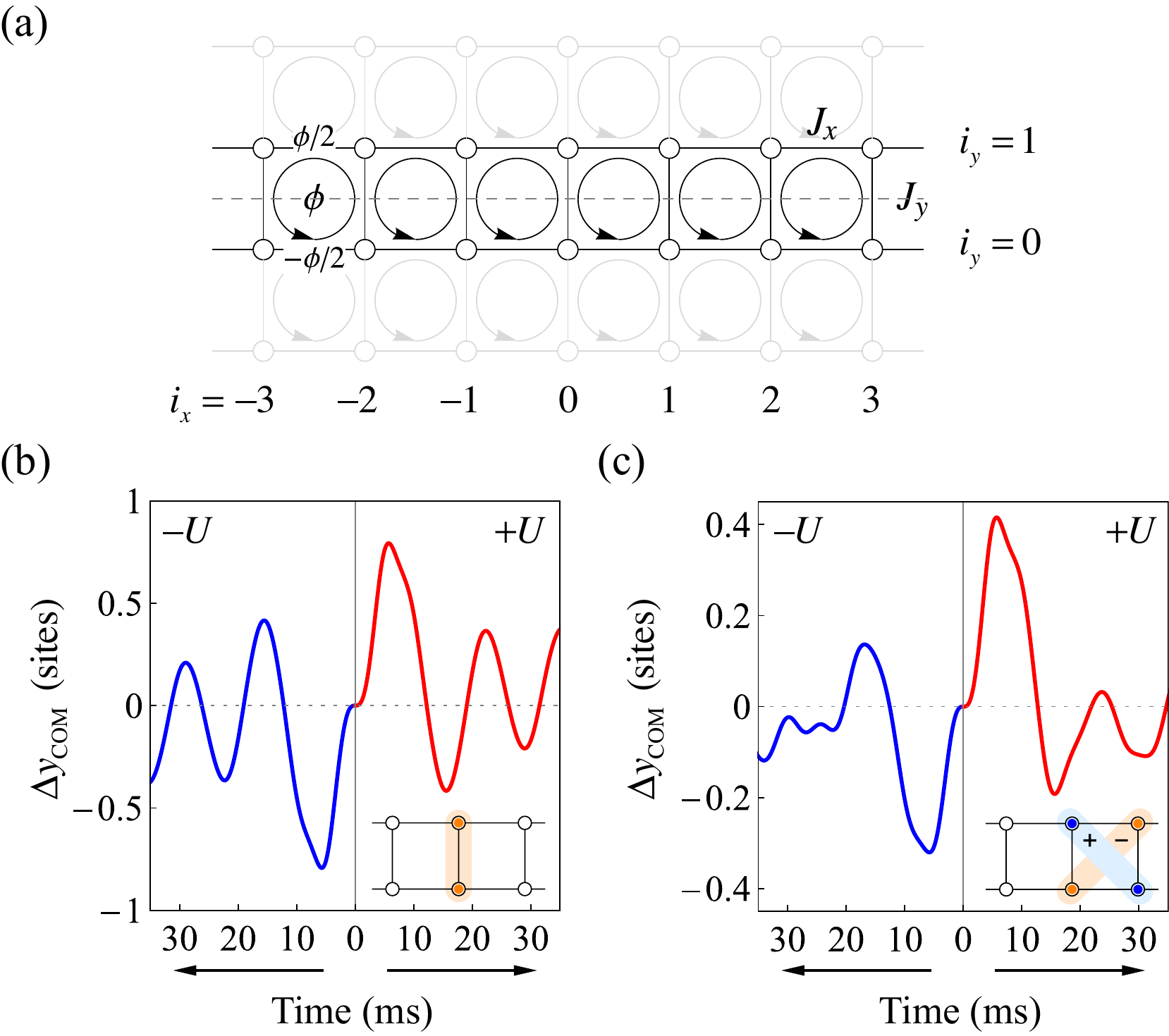}\\
  \caption{Dynamical symmetry of the Harper-Hofstadter model. (a) A schematic illustration of the Harper-Hofstadter model. An orbital magnetic field is applied to the square lattice, giving rise to flux $\phi$ per plaquette. (b) The time-dependent shearing for the attractive (left half) and repulsive (right half) interaction cases. The initial state is a two-body state located at the central rung [Eq.~(\ref{Eq:psi_AB_in_experiment})], as shown schematically by the inset. The parameters are the same as the ones in the experiment~\cite{Harvard2017}: $\phi=0.55\pi$, $(J_x, \,J_y,\, U)=2\pi\times(11,\, 34, \,131)\,\text{Hz}$. (c) The same as (b), except that we take a different initial state [Eq.~(\ref{Eq:psi_mix_rung})] as shown schematically by the inset. }\label{Fig2}
\end{figure}

\textit{Example 2:} In this case we consider atoms hopping in a square lattice with a uniform magnetic flux $\phi$ at each plaquette~\cite{Harper1955, Hofstadter1976,Bloch2013,Ketterle2013,Harvard2017}, as shown in Fig.~\ref{Fig2}(a). Here we can choose a particular gauge such that the hopping along the ${x}$ direction acquires a nontrivial phase, and the corresponding Harper-Hofstdter Hamiltonian can be written as~\cite{Harper1955, Hofstadter1976}
\begin{align}
\hat{H}_0=&-J_x\sum\limits_{i,\sigma}\left[e^{i(i_y-1/2)\phi}\hat{c}^\dag_{i_x,i_y,\sigma}\hat{c}_{i_x-1,i_y,\sigma}+\text{h.c.}\right]\nonumber\\
&-J_y\sum\limits_{i,\sigma}\left[\hat{c}^\dag_{i_x,i_y,\sigma}\hat{c}_{i_x,i_y-1,\sigma}+\text{h.c.}\right].
\end{align}
Because the time-reversal operation will change the flux $\phi$ to $-\phi$, the choice of $\hat{W}$ operator as Eqs. (\ref{W1a}) and (\ref{W1b}) does not work here. Instead, one has to include a reflection into the definition of $\hat{W}$ and the reflection axes is a middle line between $i_x=0$ and $i_x=1$, as shown by the dashed line in Fig.~\ref{Fig2}(a). That is to say, $\hat{W}$ is defined as
\begin{align}
&\hat{W}^{-1}\hat{c}_{i_x,i_y,\sigma}\hat{W}=(-1)^{i_x+i_y}\hat{c}_{i_x,1-i_y,\sigma},  \label{W2} 
\end{align}
with which $\hat{S}$ satisfies condition (i). 

In the \textbf{Harvard 2017} experiment~\cite{Harvard2017}, they consider a two-leg ladder (with $i_y=0$ and $i_y=1$) loaded with bosons. Their initial state is prepared as
\begin{align} \label{Eq:psi_AB_in_experiment}
|\Psi_0\rangle&=\frac{1}{4}\left[(\hat{c}^\dag_{0,0}+\hat{c}^\dag_{0,1})^2-(\hat{c}^\dag_{0,0}-\hat{c}^\dag_{0,1})^2\right]|0\rangle\nonumber\\
&=\hat{c}^\dag_{0,0}\hat{c}^\dag_{0,1}|0\rangle.
\end{align}
This initial state is invariant with respect to the refection defined above, and consequently, is invariant under $\hat{S}$. Furthermore, the chirality they considered is whether the atoms moving to the right is more concentrated in the upper ladder than the atoms moving to the left. To quantify the amount of chirality, they define the shearing $\Delta y_{\text{COM}}$ [i.e., the difference between the center-of-mass (COM) displacements along the $y$ direction for the right and left halves] as follows:
\begin{align}
\Delta y_{\text{COM}}(t)=\frac{\langle \hat{O}^R_{-}(t)\rangle}{\langle \hat{O}^R_{+}(t)\rangle}-\frac{\langle \hat{O}^L_{-}(t)\rangle}{\langle \hat{O}^L_{+}(t)\rangle},
\end{align}
where 
\begin{align}
&\hat{O}^R_{\pm}=\sum\limits_{i_x>0}(\hat{n}_{i_x,i_y=1}\pm \hat{n}_{i_x,i_y=0}), \\
&\hat{O}^L_{\pm}=\sum\limits_{i_x<0}(\hat{n}_{i_x,i_y=1}\pm \hat{n}_{i_x,i_y=0}). 
\end{align}
It is straightforward to show that
\begin{align}
\hat{S}^{-1}\hat{O}^{R/L}_{\pm}\hat{S}=\pm \hat{O}^{R/L}_{\pm},
\end{align}
and as a consequence of our theorem, we have
\begin{align}
\Delta {y_{{\text{COM}}}}(t)\left| {_{ + U}} \right. =  - \Delta {y_{{\text{COM}}}}(t)\left| {_{ - U}} \right. .
\end{align}
Because $\Delta {y_{{\text{COM}}}}$ is an odd function in $U$, it must be zero for all time when $U=0$. This leads to the conclusion that the charity vanishes for the non-interacting case.

The insight from this theorem is that this conclusion essentially depends on the choice of the initial state. Our theorem does not hold if the initial state does not respect the symmetry defined in Eq. (\ref{W2}), for instance, we can consider an alternative two-body state  
\begin{align} \label{Eq:psi_mix_rung}
|\Psi_0\rangle&=\frac{1}{2}(\hat{c}^\dag_{0,1}+\hat{c}^\dag_{1,0})(\hat{c}^\dag_{1,1}-\hat{c}^\dag_{0,0})|0\rangle,
\end{align}
and we shall change the summation in the definition of $\hat{O}^R_{\pm}$ to $i>1$. It is easy to show that this initial state does not respect the symmetry operation $\hat{S}$, and $\Delta {y_{{\text{COM}}}}$ is no longer an odd function in $U$. Thus, $\Delta {y_{{\text{COM}}}}$ is finite for the non-interacting case. In other word, in order to have the phenomenon of ``interaction induced chirality" observed in the \textbf{Harvard 2017} experiment~\cite{Harvard2017}, one condition is that the initial state is chosen to respect this symmetry operation $\hat{S}$ that includes reflection.  In Fig.~\ref{Fig2}, we show the numerical results for the time evolution of two bosons with these two different initial states, respectively, and the results are fully consistent with the conclusion.

\begin{figure}[t]
  \includegraphics[width=\columnwidth]{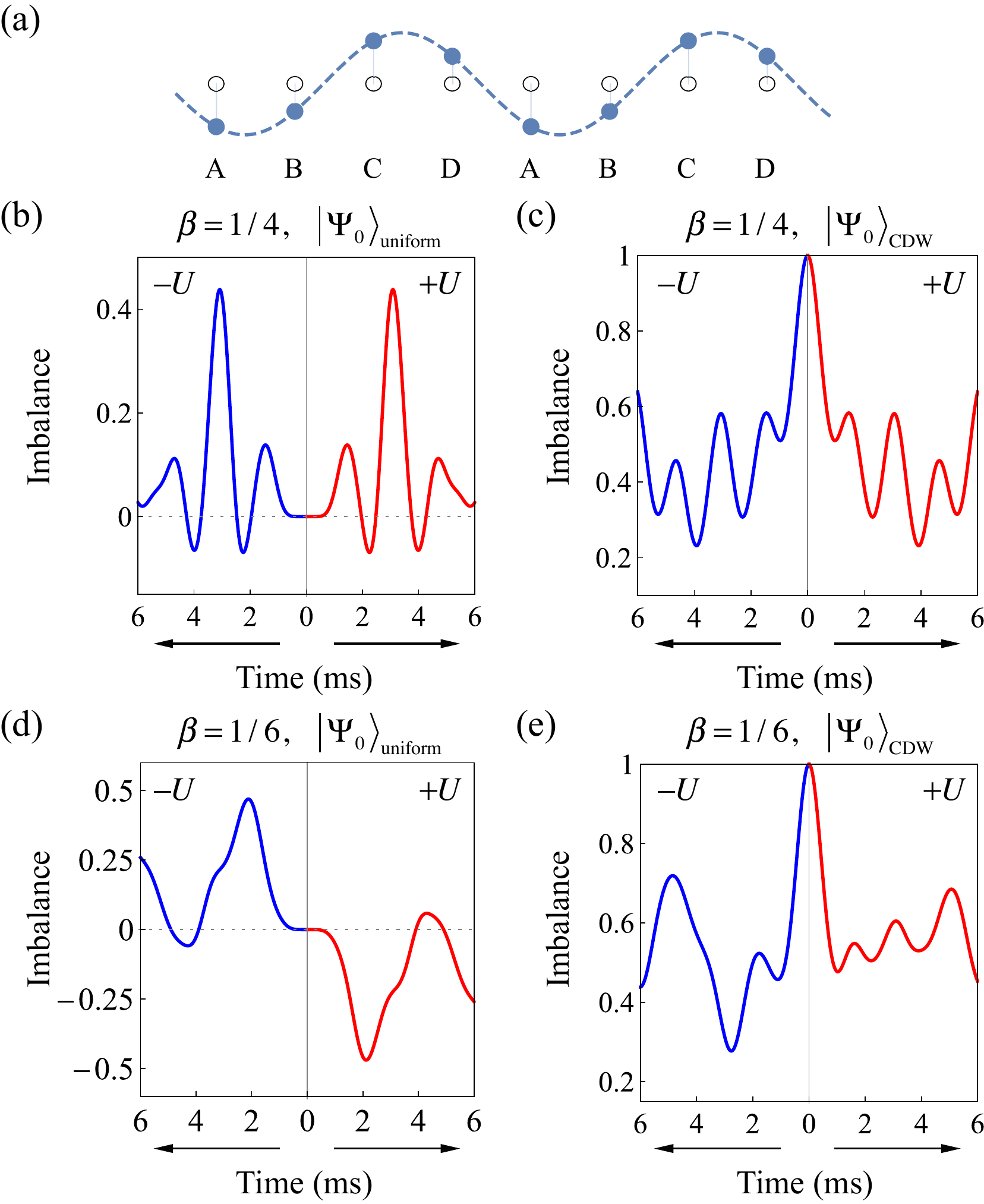}\\
  \caption{Dynamical symmetry of the fermionic Aubry-Andr{\'e} model. (a) The dashed line shows the superlattice potential $V(x)=\Delta\cos(2\pi\beta x + \theta)$ with $\beta=1/4$ and $\theta=1.0$. The open circles indicate the equilibrium positions of the particles in the absence of $\Delta$. When $\Delta\neq0$, the equilibrium positions are shifted, as shown by the filled circles. A unit cell for this case consists of four lattice sites, which are denoted as $A$, $B$, $C$, and $D$, respectively.  (b-d) The time-dependent imbalance for different lattice periodicities ($\beta=1/4$ and $1/6$) and initial states [Eqs.~(\ref{Eq.psi_uniform}) and (\ref{Eq.psi_CDW})]. The other parameters in (b-d) are the same: $\theta=1.0$, and $(J,\,\Delta,\,U)=2\pi\times(100, \,200, \,500)\,\text{Hz}$. }\label{Fig3}
\end{figure}

\textit{Example 3.} In this example we consider a one-dimensional model with an extra on-site potential energy, as schematically shown in Fig.~\ref{Fig3}(a), for which the single-particle Hamiltonian takes the form of the Aubry-Andr\'e model~\cite{AAModel1980}
\begin{align}
\hat{H}_0=\sum\limits_{i,\sigma}\left[-J(\hat{c}^\dag_{i,\sigma}\hat{c}_{i+1,\sigma}+\text{h.c.})+\Delta\cos(2\pi\beta i+\theta)\hat{n}_{i,\sigma}\right].
\end{align}
Here, $\Delta$ is the strength of a superlattice potential, $\beta$ controls the  superlattice periodicity, and $\theta$ is a phase offset. Here we consider the case that $\beta=p/q$ is a rational number. Now we discuss the following three different situations:

(i) $p$ is odd, $q$ is even and $q/2$ is also an even integer. In order for the on-site energy term to acquire a minus sign under the symmetry operation, we have to introduce a proper translational operator into the definition of $\hat{W}$, that is,
\begin{align}
\hat{W}^{-1}\hat{c}_{i,\sigma}\hat{W}=(-1)^{i}\hat{c}_{i+q/2,\sigma}.  \label{W3} 
\end{align}
If the initial state is a uniform state, it is invariant under this translation. While in the \textbf {Munich 2015} experiment~\cite{Munich2015}, they consider a CDW initial state where the density varies alternatively between even and odd sites. Since $q/2$ is also an even number, the CDW state is also invariant under this translation. They examine the time evolution of the density imbalance between the even and odd site with an operator defined as
\begin{equation}
\hat{\mathcal{I}}=\frac{\sum_{i\in \text{even}}\hat{n}_{i\sigma}-\sum_{i\in \text{odd}}\hat{n}_{i\sigma}}{\sum_{i\in \text{even}}\hat{n}_{i\sigma}+\sum_{i\in \text{odd}}\hat{n}_{i\sigma}}.
\end{equation}
In this case, we have $\hat{S}^{-1}\hat{\mathcal{I}}\hat{S}=\hat{\mathcal{I}}$. Thus, we conclude that for both the uniform and CDW states, $\langle\hat{\mathcal{I}}(t)\rangle_{+U}=\langle\hat{\mathcal{I}}(t)\rangle_{-U}$. In the \textbf {Munich 2015} experiment, $\beta=532/738\approx0.721$ which to certain extent can be reasonably approximated by $3/4$.

In Figs.~\ref{Fig3}(b) and \ref{Fig3}(c), we illustrate this with a numerical solution of two spin-$1/2$ fermionic atoms case and $\beta=1/4$. The initial state for the uniform and CDW cases are respectively taken as
\begin{align}
&|\Psi_0\rangle_{\text{uniform}}=\frac{1}{\sqrt{N}}\sum\limits_{i=1}^{N}\hat{c}^\dag_{i\uparrow}\hat{c}^\dag_{i\downarrow}|0\rangle, \label{Eq.psi_uniform}\\
&|\Psi_0\rangle_{\text{CDW}}=\frac{1}{\sqrt{N/2}}\sum\limits_{i\in \text{even}}\hat{c}^\dag_{i\uparrow}\hat{c}^\dag_{i\downarrow}|0\rangle. \label{Eq.psi_CDW}
\end{align}
We find that, for both cases, the time-dependent imbalance $\langle\hat{\mathcal{I}}(t)\rangle$ is even in $U$.

(ii) $p$ is odd, $q$ is even but $q/2$ is an odd integer. In this case, the symmetry operator for $\hat{H}_0$ should still defined as Eq. (\ref{W3}), and a uniform initial state is still invariant under this translation. Nevertheless, since $q/2$ is now odd, a CDW state defined above does not obey this symmetry. Moreover, in this case, $\hat{S}^{-1}\hat{\mathcal{I}}\hat{S}=-\hat{\mathcal{I}}$. Thus, we can conclude that, if the initial state is a uniform state, $\langle\hat{\mathcal{I}}(t)\rangle$ is odd in $U$; and if the initial state is a CDW state, there is no symmetry between positive and negative $U$. Our numerical calculation for the two-atom case with $\beta=1/6$ also confirm this conclusion, as shown in Figs.~\ref{Fig3}(d) and \ref{Fig3}(e).

(iii) $q$ is odd. In this case, no matter $p$ is even or odd, it can be shown that there is no symmetry operator can satisfy $\hat{S}^{-1}\hat{H}_0\hat{S}=-\hat{H}_0$. Therefore, there is no dynamical symmetry for this case with both the uniform and CDW initial states. 

\textit{Concluding Remark.} Our theorem provides one of rare theoretical results for the dynamics in interacting quantum many-body systems that is mathematically rigorous, universal and directly related to experiments. This result reveals profound connection between the symmetry of the single particle Hamiltonian and the interaction induced dynamics. Our results not only explain three different seemingly disparate experiments, but also offer controllable comparative examples that can be verified by future experiments. Our results may also find their usage in future cold atom experiments on the dynamics in the optical lattices, as well as the strongly correlated solid-state materials. 

\textit{Acknowledgment.} This work is supported by MOST under Grant No. 2016YFA0301600 and NSFC Grant No. 11325418 and No. 11734010.


\begin{thebibliography}{99}
\bibitem{footnote}
In condensed matter physics, the terminology of ``symmetry protected topological phase " means the existence of certain symmetry can guarantee topological classification and topological nontrivial phases. Following the same spirits, we introduce the terminology of ``symmetry protected dynamical symmetry" which indicates that the existence of certain symmetry of the static Hamiltonian can guarantee the dynamical symmetry during time evolution.
\end{thebibliography}
\end{document}